\documentclass[runningheads,a4paper]{llncs}

\usepackage[colorlinks=true,%
            linkcolor=blue,%
            citecolor=blue,%
            urlcolor=blue]{hyperref}   
\usepackage[dvips]{graphics,color} 
\usepackage{epsfig}
\usepackage[all]{xy}
\usepackage{latexsym,amssymb}
\usepackage{fancybox,epic}
\usepackage{amsmath}
\usepackage{xspace}

\usepackage{alltt}

\newcommand{\short}[1]{}

\newcommand{\clp}{CLP\xspace}

\newcommand{\sicstus}{\texttt{Sicstus}\xspace}

\newcommand{\tdg}{TDG\xspace} 
\newcommand{\bck}{block-count$(k)$\xspace}

\newcommand{\divpoints}{{\sf div\_points}}
\newcommand{\convpoints}{{\sf conv\_points}}

\begin{document}

\title{On the Generation of Test Data for Prolog by Partial Evaluation}


\author{Miguel G\'omez-Zamalloa\inst{1} \and Elvira Albert\inst{1}
  \and Germ\'an Puebla\inst{2}}

\institute{DSIC, Complutense University of Madrid, E-28040 Madrid, Spain\\
  \and CLIP, Technical University of Madrid, E-28660 Boadilla del
  Monte, Madrid, Spain}

\tocauthor{Miguel G\'omez-Zamalloa, Elvira Albert, Germ\'an Puebla}

\maketitle
\setcounter{page}{26}


\begin{abstract}
  In recent work, we have proposed an approach to Test Data Generation
  (\tdg) of imperative bytecode by \emph{partial evaluation} (PE) of
  CLP which consists in two phases: (1) the bytecode program is first
  transformed into an equivalent CLP program by means of interpretive
  compilation by PE, (2) a second PE is performed in order to
  supervise the generation of test-cases by execution of the CLP
  decompiled program. The main advantages of \tdg by PE include
  flexibility to handle new coverage criteria, the possibility to
  obtain test-case generators and its simplicity to be implemented.
  The approach in principle can be directly applied for \tdg of any
  imperative language. However, when one tries to apply it to a
  declarative language like Prolog, we have found as a main difficulty
  the generation of test-cases which cover the more complex control
  flow of Prolog. Essentially, the problem is that an intrinsic
  feature of PE is that it only computes non-failing derivations while
  in \tdg for Prolog it is essential to generate test-cases associated
  to failing computations.  Basically, we propose to transform the
  original Prolog program into an equivalent Prolog program with
  \emph{explicit failure} by partially evaluating a Prolog interpreter
  which captures failing derivations w.r.t.\ the input
  program. Another issue that we discuss in the paper is that, while
  in the case of bytecode the underlying constraint domain only
  manipulates integers, in Prolog it should properly handle the
  symbolic data manipulated by the program.
  The resulting scheme is of interest for bringing the advantages
  which are inherent in \tdg by PE to the field of logic programming.

\end{abstract}



\section{Introduction}\label{sec:intro}

Test data generation (\tdg) 
aims at automatically generating test-cases for interesting test
\emph{coverage criteria}.  The coverage criteria measure how well the
program is exercised by a test suite. Examples of coverage criteria
are: \emph{statement coverage} which requires that each line of the
code is executed; \emph{path coverage} which requires that every
possible trace through a given part of the code is executed; etc.
There are a wide variety of approaches to \tdg (see \cite{267590} for
a survey).  Our work focuses on \emph{glass-box} testing, where
test-cases are obtained from the concrete program in contrast to
\emph{black-box} testing, where they are deduced from a specification
of the program.  Also, our focus is on \emph{static} testing, where we
assume no knowledge about the input data, in contrast to
\emph{dynamic} approaches~\cite{226158} which execute the program to
be tested for concrete input values.

The standard approach to generating test-cases statically is to
perform a \emph{symbolic} execution of the
program~\cite{Meudec01,360252,GotliebBR00}, where the contents of
variables are expressions rather than concrete values. The symbolic
execution produces a system of \emph{constraints} consisting of the
conditions to execute the different paths. This happens, for instance,
in branching instructions, like if-then-else, where we might want to
generate test-cases for the two alternative branches and hence
accumulate the conditions for each path as constraints. The symbolic
execution approach is usually combined with the use of
\emph{constraint solvers} in order to: handle the constraints systems
by solving the feasibility of paths and, afterwards, to instantiate
the input variables.

TDG for declarative languages has received comparatively less
attention than for imperative languages.  In general, declarative
languages pose different problems to testing related to their own
execution models, like laziness in functional programming (FP) and
failing derivations in constraint logic programming (CLP). The
majority of existing tools for FP are based on black-box testing (see
e.g.~\cite{ClaessenH00}). An exception is
\cite{DBLP:conf/ppdp/FischerK07} where a glass-box testing approach is
proposed to generate test-cases for Curry.  In the case of \clp,
test-cases are obtained for Prolog
in~\cite{LuoBoyer1992,issta/BelliJ93,aplas/ZhaoGQC07}; and very
recently for Mercury in~\cite{DegraveVanhoof-LOPSTR08}. Basically the
test-cases are obtained by first computing constraints on the input
arguments that correspond to execution paths of logic programs and
then solving these constraints to obtain test inputs for such paths.

In recent work \cite{tdg-lopstr08}, we have proposed to employ
existing \emph{partial evaluation} (PE) techniques developed for \clp
in order to automatically generate \emph{test-case generators} for
glass-box testing of bytecode.
PE \cite{pevalbook93} is an automatic program transformation technique
which has been traditionally used to specialise programs w.r.t. a
known part of its input data and, as Futamura predicted, can also be
used to compile programs in a (source) language to another (object)
language (see \cite{Futamura:71:54}).  The approach to \tdg by PE of
\cite{tdg-lopstr08} consists of two independent CLP PE phases. (1)
First, the bytecode is transformed into an equivalent (decompiled) CLP
program by specialising a bytecode interpreter by means of existing PE
techniques.  (2) A second PE is performed in order to supervise the
generation of test-cases by execution of the CLP decompiled program.
Interestingly, it is possible to employ control strategies previously
defined in the context of CLP PE in order to capture \emph{coverage
  criteria} for glass-box testing of bytecode.  A unique feature of
this approach is that, this second PE phase allows generating not only
test-cases but also test-case \emph{generators}.  Another important
advantage is that, in contrast to previous work to \tdg of bytecode,
it does not require devising a dedicated symbolic virtual machine.

In this work, we study the application of the above approach to \tdg
by means of PE to the Prolog language. Compared to \tdg of an
imperative language \cite{tdg-lopstr08}, dealing with Prolog brings in
as the main difficulty to generate test-cases associated to failing
computations. This happens because an intrinsic feature of PE is that
it only produces results associated to the \emph{non-failing}
derivations. While this is what we need for \tdg of an imperative
language (like bytecode above), we now want to capture non-failing
derivations in Prolog and still rely on a standard partial
evaluator. Our proposal is to transform the original Prolog program
into an equivalent Prolog program with explicit failure by partially
evaluating a Prolog interpreter which captures failing derivations
w.r.t.\ the input program.  This transformation is done in the phase
(1) above. As another difference, in the case of bytecode, the
underlying constraint domain only manipulates integers.  However, the
above phase (2) should properly handle the data manipulated by the
program in the case of Prolog.
Compared to existing approaches to \tdg of
Prolog~\cite{issta/BelliJ93,LuoBoyer1992}, our approach basically is
of interest for bringing the advantages which are inherent in \tdg by
PE to the field of Prolog:
\begin{itemize}
\item[(i)] It is \emph{more powerful} in that we can produce test-case
  generators which are CLP programs whose execution in CLP returns
  further test-cases on demand without the need to start the \tdg
  process from scratch; \item[(ii)] It is more \emph{flexible}, as
  different coverage criteria can be easily incorporated to our
  framework just by adding the appropriate local control to the
  partial evaluator.  \item[(iii)] It is \emph{simpler} to implement
  compared to the development of a dedicated test-case generator, as
  long as a CLP partial evaluator is available.
\end{itemize}

The rest of the paper is organized as follows. In the next section, we
give some basics on PE of logic programs and describe in detail the
approach to \tdg by PE proposed in
\cite{tdg-lopstr08}. Sect.~\ref{sec:control_flow} discusses some
fundamental issues like the Prolog control-flow and the notion of
computation path. Then, Sect.~\ref{sec:explicit_failure} describes the
program transformation to make failure explicit,
Sect.~\ref{sec:gener-test-cases} outlines existing methods to properly
handle symbolic data during the TDG phase, and finally
Sect.~\ref{sec:future} concludes and discusses some ideas for future
work.



\section{Basics of TDG by Partial Evaluation}\label{sec:basics-tdg-partial}

\begin{figure}[t]
\begin{minipage}{\textwidth}\scriptsize
\begin{center}
\includegraphics[scale=0.78]{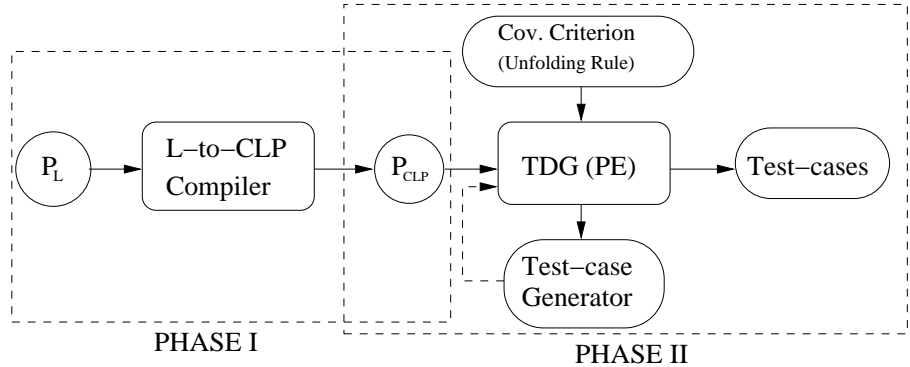}
\end{center}
\end{minipage}
\caption{General scheme of TDG by Partial Evaluation of CLP}\label{fig:overview}
\end{figure}
\newcommand{\unfold}{{\sf unfold}}
\newcommand{\abstraction}{{\sf abstract}}
\newcommand{\unfoldT}{{\sf unfold}_{\embt}}
\newcommand{\abstractionT}{{\sf abstract}_{\embt}}
\newcommand{\unfoldTA}{{\sf unfold}_{\embt}^{\cal A}}
\newcommand{\abstractionTA}{{\sf abstract}_{\embt}^{\lambda}}
\newcommand{\T}{T^{pe}\xspace} 
\newcommand{\leaves}{\emph{leaves}\xspace}
\newcommand{\codegen}{{\sf codegen}}

In this section we recall the basics of partial evaluation of logic
programming and summarize the general approach of relying on partial
evaluation of CLP for TDG of an imperative language, as proposed in
\cite{tdg-lopstr08}.

\subsection{Partial Evaluation and its Application to Compilation}\label{sec:pe-basics}

We assume familiarity with basic notions of logic programming and
partial evaluation (see e.g.~\cite{gallagher:pepm93-short}).
Partial evaluation is a semantics-based program transformation
technique which specialises a program w.r.t.\ given input data, hence,
it is often called \emph{program specialisation}.  Essentially,
partial evaluators are non-standard interpreters which evaluate goals
as long as termination is guaranteed and specialisation is considered
profitable.  In logic programming, the underlying technique is to
construct (possibly) \emph{incomplete} SLD trees for the set of atoms
to be specialised.  In an incomplete tree, it is possible to choose
\emph{not} to further unfold a goal. Therefore, the tree may contain
three kinds of leaves: failure nodes, success nodes (which contain the
empty goal), and non-empty goals which are not further unfolded.  The
latter are required in order to guarantee termination of the partial
evaluation process, since the SLD being built may be infinite.  Even
if the SLD trees for fully instantiated initial atoms (as regards the
\emph{input} arguments) are finite, the SLD trees produced for
partially instantiated initial atoms may be infinite. This is because
the SLD for partially instantiated atoms can have (infinitely many)
more branches than the actual SLD tree at run-time.

The role of the \emph{local control} is to determine how to construct
the (incomplete) SLD trees. In particular, the \emph{unfolding rule}
decides, for each resolvent, whether to stop unfolding or to continue
unfolding it and, if so, which atom to select from the resolvent.  On
the other hand, partial evaluators need to compute SLD-trees for a
number of atoms in order to ensure that all atoms which appear in
non-failing leaves of incomplete SLD trees are ``covered'' by the root
of some tree (this is known as the closedness condition of partial
evaluation \cite{gallagher:pepm93-short}).  The role of the
\emph{global control} is to ensure that we do not try to compute SLD
trees for an infinite number of atoms.  The usual way of achieving
this is by applying an \emph{abstraction operator} which performs
``generalizations'' on the atoms for which SLD trees are to be built.
The global control returns a set of atoms $T$. Finally, the partial
evaluation can then be systematically extracted from the set $T$ (see
\cite{gallagher:pepm93-short} for details).

Traditionally, there have been two different approaches regarding the
way in which control decisions are taken, \emph{on-line} and
\emph{off-line} approaches. In \emph{online} PE, all control decisions
are dynamically taken during the specialisation phase. In
\emph{offline} PE, a set of previously computed annotations (often
manually provided) gives information to the control operators to
decide, 1) when to stop unfolding (\emph{memoise}) in the local
control, and 2) how to perform generalizations in the global control.

The development of PE techniques has allowed the so-called
``interpretative approach'' to compilation which consists in
specialising an interpreter w.r.t.\ a fixed object code.  Interpretive
compilation was proposed in Futamura's seminal work
\cite{Futamura:71:54}, whereby compilation of a program $P$ written in
a (\emph{source}) programming language $L_S$ into another
(\emph{object}) programming language $L_O$ is achieved by partially
evaluating an interpreter for $L_S$ written in $L_O$
w.r.t.\ $P$.  The advantages of interpretive (de-)compilation w.r.t.\
dedicated (de-)compilers are well-known and discussed in the PE
literature (see, e.g., \cite{jvm-pe-padl07-short}). Very briefly, they
include: \emph{flexibility}, it is easier to modify the interpreter in
order to tune the decompilation (e.g., observe new properties of
interest); \emph{easier to trust}, it is more difficult to prove
that ad-hoc decompilers preserve the program semantics;
  \emph{easier to maintain}, new changes in the language semantics can
  be easily reflected in the interpreter.

\subsection{A General Scheme to TDG of Imperative Languages by PE}\label{sec:general-scheme}

In recent work, we have proposed an approach to Test Data Generation
(\tdg) by PE of CLP \cite{tdg-lopstr08} and used it for \tdg of
bytecode.  The approach is generic in that the same techniques can be
applied to TDG other both low and high-level imperative languages. In
Figure~\ref{fig:overview} we overview the main two phases of this
technique.  In {\bf Phase I}, the input program written in some
(imperative) language $L$ is compiled into an equivalent CLP program
$P_{CLP}$.  This compilation can be achieved by means of an ad-hoc
decompiler (e.g., an ad-hoc decompiler of bytecode to Prolog
\cite{decomp-oo-prolog-lopstr07}) or, more interestingly, can be
achieved automatically by relying on the first Futamura projection by
means of PE for logic programs as explained above (e.g.,
\cite{HGScam06,jvm-pe-padl07-short,mod-decomp-scam08}).

Now, the aim of {\bf Phase II} is to generate test-cases which
traverse as many different execution paths of $P_L$ as possible,
according to a given coverage criteria. From this perspective,
different test data will correspond to different execution paths. With
this aim, rather than executing the program starting from different
input values, the standard approach consists in performing
\emph{symbolic execution} such that a single symbolic run captures the
behavior of (infinitely) many input values.
The central idea in symbolic execution is to use constraint variables
instead of actual input values and to capture the effects of
computation using constraints. Hence, the compilation from $L$ to CLP
allows us to use the standard CLP execution mechanism to carry out
this phase. In particular, by running the $P_{CLP}$ program without
input values, each successful execution corresponds to a different
computation path in $P_L$.

Rather than relying on the standard execution mechanism, we have
proposed in \cite{tdg-lopstr08} to use PE of CLP to carry out {\bf
  Phase II}.  Essentially, we can rely on a CLP partial evaluator
which is able to solve the constraint system, in much the same way as
a symbolic abstract machine would do.  Note that performing symbolic
execution for \tdg consists in building a finite (possibly unfinished)
evaluation tree by using a non-standard execution strategy which
ensures both a certain coverage criterion and termination.
This is exactly the problem that \emph{unfolding rules}, used in
partial evaluators of (C)LP, solve.  In essence, partial evaluators
are non-standard interpreters which receive a set of partially
instantiated atoms and evaluate them as determined by the so-called
unfolding rule. Thus, the role of the unfolding rule is to supervise
the process of building finite (possibly unfinished) SLD trees for the
atoms. This view of \tdg as a PE problem has important
advantages. First, we can directly apply existing, powerful, unfolding
rules developed in the context of PE. Second, it is possible to
explore additional abilities of partial evaluators in the context of
\tdg. In particular, the generation of a residual program from the
evaluation tree returns a program which can be used as a
\emph{test-case generator}, i.e., a CLP
program whose execution in CLP returns further test-cases on
demand 
without the need to start the \tdg process from scratch.
In the rest of the paper, we study the application of this general
approach to \tdg of Prolog programs.





\section{Computation Paths for Test Data Generation of Prolog }\label{sec:control_flow}

As we have already mentioned, test data generation is about producing
test-cases which traverse as many different execution paths as
possible. From this perspective, different test data should correspond
to different execution paths.  Thus, a main concern is to specify the
computation paths for which we will produce test-cases. This requires
first to determine the control flow of the considered language.  In
this section, we aim at defining the control flow of Prolog programs
that we will use for \tdg.  Test data will be generated for the
computation paths in the control flow.

\subsection{The  Control Flow of Prolog}

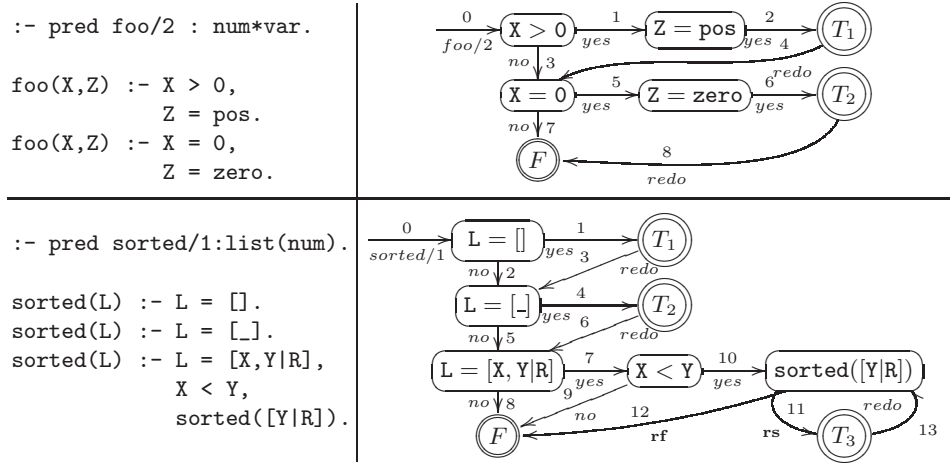
\begin{figure}[t]
\begin{center}
\begin{tabular}{c|c}
\begin{minipage}{4.5cm}\footnotesize
\begin{verbatim}
:- pred foo/2 : num*var.

foo(X,Z) :- X > 0,
            Z = pos.
foo(X,Z) :- X = 0,
            Z = zero.
\end{verbatim}
\end{minipage}
& 
\begin{minipage}{7cm}\footnotesize
\entrymodifiers={+[F-:<5pt>]}
\xymatrix@!R=0.6pt{
*{}{~~}\ar[r]_(0.3){foo/2}^(0.3)0 &
  {\tt X>0}\ar[d]^3_{no}\ar[r]^{1}_(.37){yes}
    & {\tt Z=pos}\ar[r]^{2}_(.42){yes}
    & *++[o][F=]{T_1}\ar@(dl,ur)[lld]_(.22){4}^(.22){redo} \\
*{} & {\tt X=0}\ar[d]^7_{no}\ar[r]^(.5){5}_(.37){yes}
    & {\tt Z=zero}\ar[r]^{6}_(.5){yes}
    & *++[o][F=]{T_2} \ar@(d,r)[lld]_(.65){8}^(.65){redo} \\
*{} & *++[o][F=]{F} & *{} & *{}
}
\entrymodifiers={}

\end{minipage}
\\~
\vspace{-.3cm}
\\\hline 
\begin{minipage}{4.5cm}\footnotesize
\begin{verbatim}
:- pred sorted/1:list(num).

sorted(L) :- L = [].
sorted(L) :- L = [_].
sorted(L) :- L = [X,Y|R],
             X < Y,
             sorted([Y|R]).
\end{verbatim}
\end{minipage}
& 
\begin{minipage}{7cm}\footnotesize
\vspace{.2cm}
\entrymodifiers={+[F-:<5pt>]}
\xymatrix@!R=0.6pt{
*{}{~~}\ar[r]_(0.3){sorted/1}^(0.3)0 &
  {\tt ~L=[]~}\ar[d]^2_{no}\ar[r]^{1}_(.37){yes} & 
      *++[o][F=]{T_1} \ar[ld]_(.45){3}^(.22){redo} \\
*{} & {\tt L=[\_]}\ar[d]^{5}_{no}\ar[r]^{4}_(.35){yes} & 
      *++[o][F=]{T_2} \ar[ld]_(.45)6^(.22){redo} \\
*{} & {\tt L=[X,Y|R]}\ar[r]^(.55){7}_(.55){yes}\ar[d]^{8}_{no} & 
    {\tt X < Y}\ar[r]^(.35){10}_(.35){yes}\ar[ld]_(.55){9}^(.52){no} &
    {\tt sorted([Y|R])}\ar@/^/[lld]_(.59){12}^(.55){\bf rf}\ar@(l,l)[d]^(.7){11}_(.7){\bf rs}\\
*{} & *++[o][F=]{F} & *{} & *++[o][F=]{T_3}\ar@(r,r)[u]_(.35){13}^(.35){redo}
}
\entrymodifiers={}

\end{minipage}
\end{tabular}

\end{center}
\caption{Working example. Prolog code and CFGs.}
\label{fig:cfgs}
\end{figure}

As usual a Prolog program consists of a set of predicates, where each
predicate is defined as a sequence of clauses of the form $H$ :-
$B_1,\ldots,B_m$ with $m \geq 0$. A predicate is univocally determined
by its \emph{predicate signature} $p/n$, being $p$ the name of the
predicate and $n$ its arity. Throughout the rest of the paper we will
consider Prolog programs with the following features:
\begin{itemize}
\item Rules are normalized, i.e., arguments in the head of the rule
  are distinct variables. The corresponding bindings will appear
  explicitly in the body as unifications.
\item Atoms appearing in the bodies of rules can be: unifications
  (considered as builtins), calls to defined predicates, term checking
  builtins (\texttt{==/2}, \texttt{\textbackslash==/2}, etc), and
  arithmetic builtins (\texttt{is/2}, \texttt{</2}, \texttt{=</2},
  etc). Other typical Prolog builtins like \texttt{fail/0},
  \texttt{!/0}, \texttt{if/3}, etc, have been deliberately left out to
  simplify the presentation.
\item All predicates must be moded and well-typed. We will assume the
  existence of a ``\texttt{:- pred}'' declaration associated with each
  predicate specifying the type expected for each argument (see as
  example the declarations in Fig.~\ref{fig:cfgs}). Note that this
  assumption is sensible in the context of \tdg (as the aim is the
  automatic generation of test \emph{input}).  Also, it should not be
  a limitation as analyses that can automatically infer this
  information exist.
\end{itemize}

The control flow in Prolog programs is significantly more complex than
in traditional imperative languages. The declarative semantics of
Prolog implies some additional features like: 1) several forms of
backtracking, induced by the failure of a sub-goal, or by
non-deterministic predicates; or 2) forced control flow change by the
predicate ``cut''. Traditionally, control-flow graphs (CFGs for short)
are used to statically represent the control-flow of
programs. Typically, in a CFG, nodes are blocks containing a set of
sequential instructions, and edges represent the flows that the
program can follow w.r.t. the semantics of the corresponding
programming language.  In the literature, CFGs for Prolog (and
Mercury) have been used for the aim of \tdg
in~\cite{LuoBoyer1992,aplas/ZhaoGQC07} (\cite{DegraveVanhoof-LOPSTR08}
for Mercury). In particular, CFGs determine the computation paths for
which test-cases will be produced. Our framework relies on the CFGs
of~\cite{LuoBoyer1992,aplas/ZhaoGQC07} which are known as
\emph{p-flowgraph}'s.\footnote{The difference with the CFGs
  in~\cite{LuoBoyer1992,aplas/ZhaoGQC07} is that they consider one
  additional node per clause to explicitly represent the unification
  of the head of the rule. This is not needed in our case since
  predicates are normalised.}  As will be explained later, there are
some differences between these CFGs and the ones in
\cite{DegraveVanhoof-LOPSTR08} which lead to different test-cases.

Figure~\ref{fig:cfgs} depicts the Prolog code together with the
corresponding CFGs for predicates \texttt{foo/2} and
\texttt{sorted/1}. Predicate \texttt{foo/2}, given a number in its
first argument, returns, in the second one, the value \texttt{pos} if
the number is positive and \texttt{zero} if it is zero. If the number
is negative, it just fails. Predicate \texttt{sorted/1}, given a list
of numbers, checks whether the list is strictly sorted, in that case
it succeeds, otherwise it fails. The CFGs contain the following nodes:
\begin{itemize}
\item a non-terminal node associated to each atom in the body of each
  clause,
\item a set of terminal nodes ``T$_i$'' representing the success of
  the $i$-th clause, and
\item the terminal node ``F'' to represent failure.
\end{itemize}
As regards edges, in principle all non-terminal nodes have two output
flows, corresponding to the cases where the builtin or predicate call
succeeds or fails respectively. They are labeled as ``yes'' or ``no''
for builtins (including unifications), and as ``\textbf{rs}''
(\emph{return-after-success}) or ``\textbf{rf}''
(\emph{return-after-failure}) for predicate calls. There is an
exception in the case of unifications where one of the arguments is a
variable, in which case the unification cannot fail. This can be known
statically by using the mode information. See for example nodes
``\texttt{Z=pos}'' and ``\texttt{Z=zero}'' in the \texttt{foo/2}
CFG. Both ``yes'' and ``\textbf{rs}'' edges point to the node
representing the next atom in the clause or to the corresponding
``T$_i$'' node if the atom is the last one. Finally, each ``T$_i$''
node has an output edge labeled as ``redo'' to represent the case in
which the predicate is asked for more solutions. All ``no'',
``\textbf{rf}'' and ``redo'' edges point either to the node
corresponding to the first previous non-deterministic call in the same
clause, or the first node of the following clause, or the ``F'' node
if no node meets the above conditions. See as an example the
``\textbf{rs}'' and ``\textbf{rf}'' edges from the non-terminal node
for {\tt sorted([Y|R])}.

\subsection{Generating Test Data for Computation Paths}

In order to define the computation paths determined by the CFGs, every
edge in every CFG is labeled with a unique natural number. An special
edge labeled with ``0'' and $p/n$ represents the entry of predicate
$p/n$.

\begin{definition}[Computation sub-path]
  Given the CFG for predicate $P$, a \emph{computation sub-path} is a
  sequence of numeric labels (natural numbers) $\langle
  l_1,\ldots,l_n\rangle$ s.t.: 
  \begin{itemize}
  \item $l_1$ corresponds to either an entry, an ``\textbf{rs}'', an
    ``\textbf{rf}'' or a ``redo'' edge,
  \item $l_n$ leads to a terminal node or to a predicate call, and
  \item for all consecutive labels $l_i,l_j$, there exists a node
    corresponding to a builtin in the CFG of $P$, for which $l_i$ is
    an input flow and $l_j$ is an output flow.
  \end{itemize}
\end{definition}

\begin{definition}[Computation path]
  Given the CFGs corresponding to the set of predicates defining a
  program, a \emph{computation path} (CP for short) for predicate $p$
  is a concatenation $sp_1 \cdots sp_m$ ($m \geq 1$) of computation
  sub-paths such that:
  \begin{itemize}
  \item First label in $sp_1$ is either $0$, in which case we say it
    is a \emph{full} CP, or corresponds to a ``redo'' edge, in which
    case we say it is a \emph{partial} CP (PCP for short).
  \item Last label in $sp_m$ leads to a terminal node in the CFG of
    $p$. If it is a $T$ node the CP is said to be \emph{successful}
    otherwise it is called \emph{failing}.
  \item For all $sp_k$ whose last label leads to a node corresponding
    to a predicate call, $cp = sp_{k+1} \cdots sp_{j}$, $j>k$ is a CP
    for the called predicate, and:
    \begin{itemize}
    \item if $cp$ is successful then the first label in $sp_{j+1}$
      corresponds to an ``\textbf{rs}'' edge,
    \item otherwise ($cp$ is failing), it corresponds to an
      \textbf{rf} edge.
    \end{itemize}
  \item For all $sp_k$ whose first label corresponds to a ``redo''
    edge flowing from a ``T$_a$'' node in the CFG of predicate $q$,
    $\exists sp_j$, $j<k$, whose first label corresponds either to an
    entry edge or to a ``redo'' edge flowing from ``T$_b$'', $b<a$, of
    the CFG of $q$.
  \end{itemize}
  If a CP contains at least one label corresponding to a ``redo''
  flow, then the CP is said to be an \emph{after-retry} CP. The rest
  of the CPs are \emph{first-try} CPs.
\end{definition}

\noindent
For example in \texttt{foo/2}, $p_1$=$\langle 0,1,2\rangle$ and
$p_2$=$\langle 0,3,5,6\rangle$ are first-try successful CPs;
$p_3$=$\langle 0,3,7\rangle$ is a first-try failing branch;
$p_4$=$\langle 0,1,2\rangle \cdot \langle4,5,6\rangle$ is an
after-retry successful CP (although this one is unfeasible as $X>0$
and $X=0$ are disjoint conditions), and $p_5$=$\langle 0,1,2\rangle
\cdot \langle4,7\rangle$ is an after-retry failing branch. In
\texttt{sorted/1}, $p_6$=$\langle 0,2,5,7,10\rangle \cdot \langle
0,2,4\rangle \cdot \langle 11\rangle$ is a first-try successful CP and
$p_7$=$\langle 0,2,5,7,10\rangle \cdot \langle 0,2,5,7,9\rangle \cdot
\langle 12\rangle$ is a first-try failing CP.
It is interesting to observe the correspondence between the CPs and
the test data that make the program traverse them.  In \texttt{foo/2},
$p_1$ is followed by goal \texttt{foo(1,Z)}, $p_2$ by goal
\texttt{foo(0,Z)}, $p_3$ by \texttt{foo(-1,Z)}, $p_4$ is an unfeasible
path, and $p_5$ is followed by \texttt{foo(0,Z)} when we ask for more
solutions.  As regards \texttt{sorted/1}, $p_6$ is followed by the
goal \texttt{sorted([0,1])} and $p_7$ by \texttt{sorted([0,1,0])}.  As
we will see in Sect.~\ref{sec:gener-test-cases}, these will become
part of the test-cases that we automatically infer.

A key feature of our CFGs is that they make explicit the fact that
after failing with a clause the computation has to re-try with the
following clause, unless a non-deterministic call is left
behind. E.g., in \texttt{foo/2} the CFG makes explicit that the only
way to get a first-try failing branch is through the CP
$\langle0,3,7\rangle$, hence traversing, and failing in, both
conditions $X>0$ and $X=0$. Therefore, a test data to obtain such a
behavior will be a negative number for argument $X$. Other approaches,
like the one in~\cite{DegraveVanhoof-LOPSTR08}, do not handle flows
after failure in the same way. In fact,
in~\cite{DegraveVanhoof-LOPSTR08}, edge ``3'' in \texttt{foo/2} goes
directly to node ``F''. It is not clear if these approaches are able
to obtain such a test data.
As another difference with previous approaches to \tdg of Prolog, we
want to highlight that we use CFGs just to reason about the program
transformation that will be presented in the following section and, in
particular, to clarify which features we want to capture. However, in
previous approaches, test-cases are deduced directly from the CFGs.



\section{A Program Transformation to Make Failure Explicit}\label{sec:explicit_failure}

As we outlined in Sect.~\ref{sec:intro}, an intrinsic feature of the
second phase of our approach is that it can only produce results
associated to non-failing derivations.  This is the main reason why
the general approach to TDG by PE sketched in
Sect.~\ref{sec:basics-tdg-partial} is directly applicable only to TDG
of imperative languages.  To enable its application to Prolog, we
propose a program transformation which makes failure explicit in the
Prolog program.  The specialisation of meta-programs has been proved
to have a large number of interesting
applications\cite{gallagher:pepm93-short}.  Futamura projection's to
derive compiled code, compilers and compiler generators fall into this
category. The specialization of meta-interpreters for non-standard
computation rules has also been studied. Furthermore, language
extensions and enhancements can be easily expressed as
meta-interpreters which perform additional operations to the standard
computation. In short, program specialisation offers a general
compilation technique for the wide variety of procedural
interpretations of logic programs.  Among them, we propose to carry
out our transformation which makes failure in logic programs explicit
by partially evaluating a Prolog meta-interpreter which captures
failing derivations w.r.t. the original program. First, in
Sect.~\ref{sec:interpreter} we describe such a meta-interpreter
emphasizing the Prolog control features which we want to capture.
Then, Sect.~\ref{sec:control_pe} describes the control strategies
which have to be used in PE in order to produce an effective
transformation.

\subsection{A Prolog Meta-Interpreter to Capture Failure}\label{sec:interpreter}

Given a Prolog program and given a goal, our aim is to define an
interpreter in which the computation of the program and goal produces
the same results as the ones obtained by using the standard Prolog
computation but with the difference that failure is never
reported. Instead, an additional argument $Answer$ will be bound to
the value ``yes'', if the computation corresponds to a successful
derivation, and to ``no'' if it corresponds to a failing derivation.
Predicate \texttt{solve/4} is the main predicate of our
meta-interpreter whose first and second arguments are the predicate
signature and arguments of the goal to be executed; and its third
argument is the answer; by now we ignore the last argument. For
instance, the call \texttt{solve(foo/2,[0,Z],Answer,\_)} succeeds with
${\tt Z=zero}$ and ${\tt Answer=yes}$, and
\texttt{solve(foo/2,[-1,Z],Answer,\_)} also succeeds, but with ${\tt
  Answer=no}$. The interpreter has to handle the following issues:

\begin{enumerate}
\item The Prolog \emph{backtracking} mechanism has to be explicitly
  implemented. To this aim, a stack of \emph{choice points} is carried
  along during the computation so that:
\begin{itemize}
\item if the derivation fails: (1) when the stack is empty, it ends up
  with success and returns the value ``no'', (2) otherwise, the
  computation is resumed from the last choice point, if any;
\item if it successfully ends: (1) when the stack is empty, the
  computation finishes with answer ``yes'', (2) otherwise, the
  computation is resumed from the last choice point.
\end{itemize}
\item When backtracking occurs, all variable bindings, between the
  current point and the choice point to resume from, have to be
  undone.
\item The interpreter has to be implemented in a \emph{big-step}
  fashion. This is a requirement for obtaining an effective
  decompilation. More details are given in Sect.~\ref{sec:control_pe}.
\end{enumerate}

\begin{figure}[t]
\begin{center}
\scriptsize
\begin{minipage}{5.4cm}
\begin{verbatim}
solve(P/Ar,Args,Answer,TNCPs) :-
   pred(P/Ar,_),
   build_s0(P/Ar,Args,S0,OutVs),
   exec(Args,S0,Sf),
   Sf = st(_,_,_,OutVs',Answer,TNCPs/_),
   OutVs' = OutVs.

exec(_,S,Sf) :- 
   S = st(_,[],[],OutVs,yes,NCPs),
   Sf = st(_,_,_,OutVs,yes,NCPs).
exec(_,S,Sf) :- 
   S = st(_,[],[_|_],OutVs,yes,NCPs),
   Sf = st(_,_,_,OutVs,yes,NCPs).
exec(_,S,Sf) :- 
   S = st(_,_,[],OutVs,no,TNCPs/0),
   Sf = st(_,_,_,OutVs,no,TNCPs/0).
exec(Args,S,Sf) :- 
   S = st(_,[],[CP|CPs],_,yes,TNCPs/0),
   build_retry_state(Args,CP,CPs,TNCPs,S'),
   exec(Args,S',Sf).
exec(Args,S,Sf) :- 
   S = st(_,_,[CP|CPs],_,no,TNCPs/0),
   build_retry_state(Args,CP,CPs,TNCPs,S'),
   exec(Args,S',Sf).
\end{verbatim}
\end{minipage}
  ~~\vline ~
\begin{minipage}{6.2cm}
\begin{verbatim}
exec(Args,S,Sf) :-
   S = st(PP,[A|As],CPs,OutVs,yes,TNCPs/ENCPs),
   PP = pp(P/Ar,ClId,Pt),
   internal(A),
   functor(A,A_f,A_ar),
   A =..[A_f|A_args],
   next(Pt,Pt'),
   solve(A_f/A_ar,A_args,Ans,ENCPs'),
   TNCPs' is TNCPs + ENCPs', 
   ENCPs'' is ENCPs + ENCPs',
   PP' = pp(P/Ar,ClId,Pt'),
   S' = st(PP',As,CPs,OutVs,Ans,TNCPs'/ENCPs''),
   exec(Args,S',Sf).
exec(Args,S,Sf) :-
   S = st(PP,[A|As],CPs,OutVs,yes,NCPs),
   PP = pp(P/Ar,ClId,Pt),
   builtin(A),
   next(Pt,Pt'),
   run_builtin(PP,A,Ans),
   PP' = pp(P/Ar,ClId,Pt'),
   S' = st(PP',As,CPs,OutVs,Ans,NCPs),
   exec(Args,S',Sf).
\end{verbatim}
\end{minipage}
\end{center}
\caption{Code of Prolog meta-interpreter to capture failure}
\label{fig:interpreter}
\end{figure}

Figure~\ref{fig:interpreter} shows an implementation of a meta-interpreter
which handles the above issues. 
The fourth argument of the main predicate \texttt{solve/4}, named
\texttt{TNCPS}, contains upon success the total number of choice
points not yet considered, whose role will be explained later. The
interpreter assumes that the program is represented as a set of
\texttt{pred/2} and \texttt{clause/3} facts. There is a
\texttt{pred/2} fact per predicate providing its predicate signature,
number of clauses and mode information; and a \texttt{clause/3} fact
per clause providing the actual code and clause identifier. Predicate
\texttt{solve/4} basically builds an initial state on \texttt{S0}, by
calling \texttt{build\_s0/4}, and then delegates on \texttt{exec/3} to
obtain the final state \texttt{Sf} of the computation. The output
information, \texttt{OutVs}, is taken from \texttt{Sf}. The state
carried along is of the form \texttt{st(PP,G,CPs,OutVs,Ans,NCPs)},
where \texttt{PP} is the current program point, \texttt{G} the current
goal, \texttt{CPs} is the stack of choice points (list of program
points), \texttt{OutVs} the list of variables in \texttt{G}
corresponding to the output parameters of the original goal,
\texttt{Ans} the current answer (``yes'' or ``no'') and \texttt{NCPs}
the number of choice points left behind. A program point is of the
form \texttt{pp(P/Ar,ClId,Pt)}, where \texttt{P/Ar}, \texttt{ClId} and
\texttt{Pt} are the predicate signature, the clause identifier and the
program point of the clause at hand. Predicate \texttt{exec/3}
implements the main loop of the interpreter. Given the current state
in its second argument it produces the final state of the computation
in the third one.  It is defined by the seven clauses which are
applied in they following situations:

\begin{description}
\item[$1^{st} cl.$] \emph{The current goal is empty, the answer
    ``yes'' and there are no pending choice points.} Then, the
  computation finishes with answer ``yes''. The current answer is
  actually used as a flag to indicate whether the previous step in the
  computation succeeded or failed (see the last two \texttt{exec/3}
  clauses).
\item[$2^{nd} cl.$] \emph{As $1^{st} cl.$ but having at least one
    choice point.} This clause represents the solution in which the
  computation ends. The 4$^{th}$ clause takes the other alternatives.
\item[$3^{rd} cl.$] \emph{The previous step failed and there are no
    pending choice points}.  Then, the computation ends with answer
  ``no''.
\item[$4^{th} cl.$] \emph{The current goal is empty, the answer
    ``yes'' and there is at least one pending choice point.} This is
  the same situation as in the $2^{nd}$ clause, however in this case
  the alternative of resuming from the last choice point is taken. The
  corresponding state \texttt{S'} is built by means of
  \texttt{build\_retry\_state/5} and the computation is resumed from
  \texttt{S'} by recursively calling \texttt{exec/3}.
\item[$5^{th} cl.$] \emph{The previous step failed and there is at
    least one pending choice point.} Then, the computation is resumed
  from the last choice point in the same way as in the previous
  clause.
\item[$6^{th} cl.$] \emph{The first atom to be solved is
    user-defined.} A call to \texttt{solve/4} handles the atom, and
  the computation proceeds with the next program point of the same
  clause which was the current one before calling
  \texttt{solve/4}. This way of solving a predicate call makes the
  interpreter \emph{big-step} (issue (3) above).
\item[$7^{th} cl.$] \emph{The first atom to be solved is a builtin.}
  Then, \texttt{run\_builtin/3} produces the corresponding answer, and
  the computation proceeds with the following program point. An
  interesting observation (also applicable for the previous clause) is
  that the answer obtained from \texttt{run\_builtin/3} (or
  \texttt{solve/4}) is now set up as the answer of the next
  state. This will make the computation go through the 3$^{rd}$ or
  5$^{th}$ clauses in the following step, if the obtained answer was
  ``no''.
\end{description}
The correspondence between these clauses and the flows in the CFGs is
as follows: clauses $1^{st}$, $2^{nd}$ and $4^{th}$ represent the
output edges from every ``T'' node. Clause $3^{rd}$ represents the
``no'' edges to ``F'' nodes and $5^{th}$ the ``no'' edges to
non-terminal nodes. Finally clauses $6^{th}$ and $7^{th}$ represents
the execution of builtins and predicate calls in non-terminal nodes
and their corresponding ``yes'' edges.

Let us now explain how the interpreter handles the above three issues.
To handle (1), a stack of choice points is carried along within the
state, initialised to contain all initial program points of each
clause defining the predicate to be solved, except for the first
one. E.g., the initial stack of choice points for \texttt{sorted/1} is
\texttt{[pp(sorted/1,2,1),pp(sorted/1,3,1)]}. How this stack is used
to perform the backtracking is already explained in the description of
the 4$^{th}$ and 5$^{th}$ \texttt{exec/3} clauses above.  As regards
issue (2), a quite simple way to implement this in Prolog is to
produce the necessary fresh variables every time the computation is
resumed. This is done inside \texttt{build\_retry\_state/5}. The
corresponding unification to link the fresh variables with the
original goal variables is made at the end (see last line of
\texttt{solve/4}). This is the reason why 1) the list of the actual
variables used in the current goal needs to be carried along within
the state; and 2) the original arguments are carried along as the
first argument of \texttt{exec/3}, as the original ground arguments
provided, have to be used when resuming from a choice point.

 Finally,
it is worth  mentioning that \texttt{solve/4} does not return
the actual stack of choice points but only the number of them. This
means that during a computation the interpreter only considers choice
points of the predicate being solved. The question is then, how can
the interpreter backtrack to the last choice point, including those
induced by other computations of \texttt{solve/4}? E.g., how can the
interpreter follow edge ``13'' in the CFG of \texttt{sorted/1}? The
interpreter performs the backtracking in the following way: 
1) The total number of
choice points left behind, \texttt{TNCPs}, is carried along within the
state and finally returned in the last argument of
\texttt{solve/4}. 2) The number of choice points corresponding to
invoked predicates, \texttt{ENCPs}, is also carried along. It is updated
right after the call to \texttt{solve/4} in the 6$^{th}$ clause of
\texttt{exec/3}. Both numbers are stored in the last argument of the
state as \texttt{TNCPs/ENCPs}. 3) Execution is resumed
from choice points of the current predicate only if ${\tt ENCPs} = 0$,
as it can be seen in the 4$^{th}$ and 5$^{th}$ clauses. Otherwise, the
computation just fails and Prolog's backtracking
mechanism is used to ask the last invoked predicate for more
solutions. This indeed means that the non-determinism of the program
is still implicit.

\subsection{Controlling  Partial Evaluation}\label{sec:control_pe}

The specialisation of interpreters has been studied in many different
contexts, see
e.g.~\cite{gallagher:pepm93-short,mod-decomp-scam08,Peralta-Gallagher-Saglam-SAS98-short}. Very
recently, \cite{mod-decomp-scam08} proposed control strategies
 to successfully specialise low-level code
interpreters w.r.t. non trivial programs. Here we demonstrate how such
guidelines can be, and should be, used in the specialisation of
non-trivial Prolog meta-interpreters. They include:
\begin{enumerate}
\item \emph{Big-step} interpreter. This solves the problem of handling
  recursion (see~\cite{mod-decomp-scam08}) and enables a compositional
  specialisation w.r.t. the program procedures (or predicates). Note
  that an effective treatment of recursion is specially important in
  Prolog programs where recursion is heavily used.
\item \emph{Optimality} issues.  Optimality must ensure that: a) the
  code to be transformed is traversed exactly once, and b) residual
  code is emitted once in the transformed program.  To achieve
  optimality, during unfolding, all atoms corresponding with
  \emph{divergence} or \emph{convergence points} in the CFG of the
  program to be transformed, has to be \emph{memoised} (see
  Sect.~\ref{sec:pe-basics}). A divergence (convergence) point is a
  program point from (to) which two or more flows originate
  (converge).
\end{enumerate}
We already explained that the interpreter in
Fig.~\ref{fig:interpreter} is big-step. As regards optimality,
by looking at the CFGs of Fig.~\ref{fig:cfgs}, we can observe: 1) all program
points are divergence points except those corresponding with
unifications in which  one argument is a variable, and 2) the first
program point of every clause, except for the one of the first clause, is a
convergence point. We assume that \textsf{conv\_points(P)} and
\textsf{div\_points(P)} denote, respectively, the set of convergence
points and divergence points of a predicate \textsf{P}. We follow the
syntax of~\cite{mod-decomp-scam08} for PE annotations. An annotation
is of the form ``$[Precond] \Rightarrow Ann ~Pred$'' where $Precond$
is an optional precondition defined as a logic formula, $Ann$ is the
kind of annotation (only \textbf{memo} in this case), and $Pred$ is a
predicate descriptor, i.e., a predicate function and distinct free
variables. Then, to achieve an effective transformation, we specialise
the interpreter in Fig.~\ref{fig:interpreter} w.r.t. the program to be
transformed by using the following annotation for each predicate
\texttt{P/Ar} in the program:
\[ {\tt PP} \in \divpoints({\tt P/Ar}) \cup \convpoints({\tt P/Ar})
\Rightarrow {\bf memo}~{\tt exec(\_,st(PP,\_,\_,\_,\_,\_),\_)}
\]
Additionally \texttt{solve/4} and \texttt{run\_builtin/3} are also
annotated to be memoised always to avoid code duplications.

This already describes how the specialisation has to be steered in the
local control. As regards the global control, the only predicate which
can introduce non-termination 
 is \texttt{exec/3}. Its first and third arguments contain a
fixed structure with variables. The second one might be problematic as it
ranges over the set of all computable states at specialisation
time. Note that the number of computable states remains finite thanks to the
big-step nature of the interpreter. Still, it can happen that
the same program point is reached with different values for the
\texttt{NCPs} sub-term of the state. Therefore, if one wants to achieve
the optimality criterion above, such argument has to be always
generalised in global control.

\begin{example}
  Figure~\ref{fig:transformed} depicts the transformed code we obtain
  for predicate
  \texttt{foo/2}. 
  It can be observed that there is a clear correspondence between the
  transformed code and the CFG in Fig.~\ref{fig:cfgs}. Thus, predicate
  \texttt{solve/4} represents the node ``\texttt{X>0}'',
  \texttt{exec\_1/5} implements its continuation, whose three clauses
  correspond to the three sub-paths $\langle3\rangle$, $\langle
  1,2\rangle$ and $\langle 1,2,4\rangle$ respectively. Predicate
  \texttt{exec\_2/4} represents the node ``\texttt{X=0}'' and
  \texttt{exec\_3/5} implements its continuation, whose two clauses
  correspond to the sub-paths $\langle 7 \rangle$ and $\langle
  5,6\rangle$. Note that edge ``8'' is not considered in the
  meta-interpreter (nor in the transformed program) as it is
  meaningless for TDG. It is worth mentioning that the transformed
  program captures the way in which variable bindings are undone. For
  instance in \texttt{solve(foo/2,[C,D],$\ldots$)}, if we keep track
  of variables \texttt{C} and \texttt{D}, it can be seen that
  \texttt{D}, which corresponds to variable \texttt{Z} in the original
  code, is only used for the final unification \texttt{F=[D]}, while
  new fresh variables are used for the unifications with \texttt{pos}
  and \texttt{zero}. However, variable \texttt{C}, which corresponds
  to variable \texttt{X} in the original code, is actually used for
  the checks in \texttt{run\_builtin\_1/2} and
  \texttt{run\_builtin\_2/2}. This turns out to be fundamental when
  trying to obtain test data associated to the \emph{first-try
    failing} CP $\langle 0,3,7\rangle$. It must be the same variable
  the one which, at the same time, is not ``$>0$'' and not
  ``=0''. Otherwise we cannot obtain a negative number as test data
  for such CP. Finally, observe that the original Prolog arithmetic
  builtins have been (automatically) transformed into their
  \texttt{clpfd} counterparts \footnote{We are using the
    \texttt{clpfd} library of \sicstus Prolog. See
    \cite{sicstus-manual-3.8} for details.}.
\end{example}

\begin{figure}[t]
\begin{center}\footnotesize

\begin{tabular}{c|c}
\begin{minipage}{6.4cm}
\begin{verbatim}
solve(foo/2,[C,D],A,B) :-
   run_builtin_1(E,C),
   exec_1(C,E,F,A,B), F = [D].

exec_1(A,no,F,G,H) :- exec_2(A,F,G,H).
exec_1(_,yes,[pos],yes,1).
exec_1(A,yes,F,G,H) :- exec_2(A,F,G,H).

exec_2(A,G,H,I) :-
   run_builtin_2(K,A), exec_3(K,G,H,I).
\end{verbatim}
\end{minipage}
\vspace{.1cm} & \vspace{.1cm}
\begin{minipage}{5.8cm}
\begin{verbatim}
exec_3(no,[_],no,0).
exec_3(yes,[zero],yes,0).

run_builtin_1(yes,A) :- A#>0.
run_builtin_1(no,A) :- \+ A#>0.

run_builtin_2(yes,A) :- A#=0.
run_builtin_2(no,A) :- \+ A#=0.
\end{verbatim}
\end{minipage}
\end{tabular}

\end{center}
\caption{Transformed code with explicit failure for {\tt foo/2}}
\label{fig:transformed}
\end{figure}



\section{Generating Test Cases by Partial Evaluation}\label{sec:gener-test-cases}

Once the original Prolog program has been transformed into an
equivalent Prolog program with explicit failure, we can use the
approach of \cite{tdg-lopstr08} to carry out \textbf{phase II}
(see Fig.~\ref{fig:overview}) and generate
test data both for  successful and failing derivations. As we have explained in
Sect.~\ref{sec:general-scheme}, the idea is to perform a second PE
over the CLP transformed program where the unfolding rule plays the
role of the coverage criterion. In~\cite{tdg-lopstr08} an unfolding
rule implementing the \emph{block-count(k)} coverage criterion was
proposed. A set of computation paths satisfies the \emph{\bck
  criterion} if it includes all terminating computation paths which
can be built in which the number of times each block is visited
 does not exceed the given $k$. The
blocks the criterion refers to are the blocks or nodes in the CFGs of
the original Prolog program. As the only form of loops in Prolog are
 recursive calls, the ``$k$'' in the \emph{\bck} actually corresponds
to the number of recursive calls which are allowed.

Unfortunately, the presence of Prolog's negation in our transformed
programs complicates this phase.  The negation will appear in the
transformed program for ``no'' branches originating from nodes
corresponding to a (possibly) failing builtin. See for example
predicates \texttt{run\_builtin\_1/3} and \texttt{run\_builtin\_2/3}
in the transformed code of \texttt{foo/2} in
Fig.~\ref{fig:transformed}. While Prolog's negation works well for
ground arguments, it gives no information for free variables, as it is
required in the evaluation performed during this TDG phase. In
particular, in the \texttt{foo/2} example, given the computation which
traverses the calls ``\texttt{\textbackslash+ A\#>0}'' and
``\texttt{\textbackslash+ A\#=0}'' (corresponding to the path $\langle
0,3,7\rangle$ in the CFG), we need to infer that
``\texttt{A$<$0}''. In other words, we need somehow to turn the
\emph{negative} information into \emph{positive} information. This
transformation is straightforward for arithmetic builtins: we just
have to replace ``\texttt{\textbackslash+ $e_1$\#=$e_2$}'' by
``\texttt{$e_1$\#\textbackslash =$e_2$}'' and
``\texttt{\textbackslash+ $e_1$\#>$e_2$}'' by
``\texttt{$e_1$\#=<$e_2$}'', etc.
\begin{example}
  This transformation allows us to obtain the following set of
  test-cases for \texttt{foo/2}:
\[
\left\{\begin{array}{ll}
\langle\mbox{\tt [1],[pos],yes/first-try}\rangle,& 
\langle\mbox{\tt [1],[\_],no/after-retry}\rangle, \\
\langle\mbox{\tt [0],[zero],yes/first-try}\rangle, &
\langle\mbox{\tt [-100],[\_],no/first-retry}\rangle
\end{array}\right\}
\]

\noindent
They correspond respectively (reading by rows) to the CPs $\langle
0,1,2\rangle$, $\langle 0,1,2\rangle \cdot \langle4,7\rangle$,
$\langle 0,3,5,6\rangle$ and $\langle 0,3,7\rangle$. Each test-case is
represented as a 3-tuple $\langle Ins,Outs,Ans\rangle$ being $Ins$ the
list of input arguments, $Outs$ the list of output arguments and $Ans$
the answer. The answer takes the form $A/B$ with $A\in\{\mbox{\tt
  yes,no}\}$ and $B\in\{\mbox{\tt
  first-try,after-retry}\}$\footnote{To simplify the presentation in
  Sect.~\ref{sec:interpreter}, we decided not include in the
  interpreter the support to calculate the
  \texttt{first-try/after-retry} value.}, so that we obtain sufficient
information about the kind of CP to which the test-case corresponds
(see Sect.~\ref{sec:control_flow}). As there are no recursive calls in
\texttt{foo/2} such test-cases are obtained using the \emph{\bck}
criterion for any $k$ (greater than $0$).  The domain used for the
integer number is $\{-100..100\}$.
\end{example}

However, it can be the case that negation involves unifications
with symbolic data. For example, the transformed code
for \texttt{sorted/1} includes the negations ``\texttt{\textbackslash+
  L=[]}'' and ``\texttt{\textbackslash+ L=[\_\textbar\_]}''. As
before,
we  might
write transformations for the negated unifications involving
lists, so that at the end it is inferred that ``\texttt{
  L=[\_,\_\textbar\_]}''. However this would be  too an ad-hoc solution
as many distinct term structures, different from lists, can appear on
 negated unifications. A solution for this problem has been
recently proposed for Mercury in the same context
\cite{DegraveVanhoof-LOPSTR08}. It roughly consists in the
following: 1) It is assumed that each predicate argument is
well-typed. 2) A domain is initialised for each variable, containing
the set of possible functors the variable can take. 3) When a negated
unification involving an output variable is found (in their
terminology a negated \emph{decomposition}), the corresponding
functor is removed from the variable domain. It is crucial at this
point the assumption that complex unifications are broken down into
simple ones. 4) Finally, a search algorithm is described to generate
particular values from the type definition and final domain for the
variable. The technique is implemented using CHR and can be directly used in
principle for our purposes as well.

On the other hand, advanced declarative languages like
TOY~\cite{LS99-TOY} make possible the co-existence of different
constraint domains. In particular, the co-existence of boolean and
numeric constraint domains enables the possibility of using
\emph{disequalities} involving both symbolic data and numbers. This
allows for example expressing the negated unifications
``\texttt{\textbackslash+ L=[]}'' and ``\texttt{\textbackslash+
  L=[\_\textbar\_]}'' as disequality constraints ``\texttt{ L/=[]}''
and ``\texttt{ L/=[\_\textbar\_]}''. Additionally, by relying on the
boolean constraint solver, the negated arithmetic builtins
``\texttt{\textbackslash+ A\#>0}'' and ``\texttt{\textbackslash+
  A\#=0}'' can be encoded as ``\texttt{(A\#>0) == false}'' and
``\texttt{(A\#=0) == false}''. This is in principle a more general
solution that we want to explore, although a thorough
experimental evaluation needs to be carried out to demonstrate its
applicability to our particular context.

\begin{example}
Now, by using any of the techniques outlined above,
 we obtain the following set of
test-cases for \texttt{sorted/1}, using \emph{block-count$(2)$} as
the coverage criterion:
\[
\left\{\begin{array}{ll}
\langle\mbox{\tt [[]],[],yes/first-try}\rangle,& 
\langle\mbox{\tt [[0]],[],yes/first-try}\rangle,\\
\langle\mbox{\tt [[0,1]],[],yes/first-try}\rangle,&
\langle\mbox{\tt [[0,1,2]],[],yes/first-try}\rangle,\\
\langle\mbox{\tt [[0,1,2,0\textbar\_]],[],no/first-try}\rangle,&
\langle\mbox{\tt [[0,1,0\textbar\_]],[],no/first-try}\rangle,\\
\langle\mbox{\tt [[0,0\textbar\_]],[],no/first-try}\rangle
\end{array}\right\}
\]
\normalsize
\noindent
They correspond respectively (reading by rows) to the CPs ``$\langle
0,1\rangle$'', ``$\langle 0,2,4\rangle$'', ``$\langle
0,2,5,7,10\rangle \cdot \langle 0,2,4\rangle \cdot \langle
11\rangle$'', ``$\langle 0,2,5,7,10 \rangle \cdot \langle 0,2,5,7,10
\rangle \cdot \langle 0,2,4\rangle \cdot \langle 11 \rangle \cdot
\langle 11\rangle$'', ``$\langle 0,2,5,7,10 \rangle \cdot \langle
0,2,5,7,10 \rangle \cdot \langle 0,2,5,7,9 \rangle \cdot \langle
12\rangle \cdot \langle 12\rangle$'', ``$\langle 0,2,5,7,10\rangle
\cdot \langle 0,2,5,7,9 \rangle \cdot \langle 12\rangle$'', ``$\langle
0,2,5,7,9\rangle$''. They are indeed all the paths that can be
followed with no more than $3$ recursive calls. This time the domain
has been set up to $\{0..100\}$.
\end{example}



\section{Conclusions and Ongoing work}\label{sec:future}

Very recently, we proposed in~\cite{tdg-lopstr08} a generic approach
to TDG by PE which in principle can be used for any imperative
language. 
However, applying this approach to TDG of a declarative language like
Prolog introduces some difficulties like the handling of failing
derivations and of symbolic data. In this work, we have sketched
solutions to overcome such difficulties. In particular, we have
proposed a program transformation, based on PE, to make failure
explicit in the Prolog programs.  To handle Prolog's negation in the
transformed programs, we have outlined existing solutions that make it
possible to turn the negative information into positive information.
Though our preliminary experiments already suggest that the approach
can be very useful to generate test-cases for Prolog, we plan to carry
out a thorough practical assessment.  This requires to cover additional
Prolog features like the module system, builtins like \texttt{cut/0},
\texttt{fail/0}, \texttt{if/3}, etc. and also to compare the results
with other TDG systems. We also want to study the integration of other
kinds of coverage criteria like \emph{data-flow} based criteria.
Finally, we would like to explore the use of static analyses in the
context of TDG. For instance, the information inferred by a
\emph{failure analysis} can be very useful to prune some of the
branches that our transformed programs have to consider.

\subsubsection*{Acknowledgments}
This work was funded in part by the Information Society Technologies
program of the European Commission, Future and Emerging Technologies
under the IST-15905 {\em MOBIUS} project, by the Spanish Ministry of
Education under the TIN-2005-09207 {\em MERIT} project, and by the
Madrid Regional Government under the S-0505/TIC/0407 \emph{PROMESAS}
project.

\bibliographystyle{plain}





\end{document}